\documentstyle[12pt]{article}

\def\a{\alpha}

\def\s{\sigma}
\def\sb{\bar{\sigma}}
\def\th{\theta}
\def\thb{\bar{\theta}}

\def\half{\frac{1}{2}}
\def\ud{\underline}

\def\G{\Gamma}

\newskip\humongous \humongous=0pt plus 1000pt minus 1000pt

\newif\ifdtup

\def\ket#1{\left| #1\right\rangle}

\def\beq{\begin{equation}}
\def\eeq{\end{equation}}

\def\beqn{\begin{eqnarray}}
\def\eeqn{\end{eqnarray}}
\relax

\def\G2{{\; \rm GeV/}c^2}
\def\G{\; \rm GeV}

\def\dotx{\dotx{\dot\overline{x}}}

\relax

\begin{document}
\begin{titlepage}
\begin{flushright}
       {\normalsize  OU-HET 299 \\  hep-th/9806139\\
           June, 1998}
\end{flushright}
%
\begin{center}
  {\large \bf Berry's Connection and $USp(2k)$ Matrix
Model }\footnote{This work is supported in part
 by the Grant-in-Aid  for Scientific Research (10640268) and
 Scientific Research Fund (97319)
from the Ministry of Education, Science and Culture, Japan.}

\vfill
         {\bf H.~Itoyama}  \\
            and \\
         {\bf T.~Matsuo}\\
        Department of Physics,\\
        Graduate School of Science, Osaka University,\\
        Toyonaka, Osaka, 560 Japan\\
\end{center}
\vfill
\begin{abstract}
 Berry's connection is computed in the $USp(2k)$ matrix model. In  
$T$ dualized quantum mechanics, the Berry phase exhibits 
a residual interaction taking place at a distance $m_{(f)}$ from the
orientifold surface via the integration of the fermions in the
fundamental representation. This is interpreted as a coupling of
the magnetic $D2$ with the electric $D4$ branes. We make a comment
on the Berry phase associated with the $6D$ nonabelian gauge anomaly
whose cancellation selects the number of flavours $n_{f}=16$.

\end{abstract}
\vfill
\end{titlepage}

Matrix models have recently received much attention as a candidate to
provide a nonperturbative formulation to superstrings and $M$ theory.
The models proposed so far include the original one for $M$ theory
\cite{BFSS} and its heterotic counterpart \cite{Mhetero} as well as the
$IIB$ matrix model \cite{IKKT} in zero dimension. This latter one may
be referred to as reduced model. In ref. \cite{IT1}, the reduced model
descending from Type $I$ superstrings based on $USp$ Lie algebra has
been found and its connection to $F$ theory \cite{Vafa} has been
suggested. Physical implications and identifications of the $USp(2k)$
matrix model have been further explored in \cite{IT2}. It is a matrix
model which embodies Type $I$ superstrings as that on
$T^{6}/{\cal Z}^{2}$ orientifold. The possible choice of the model has
been found to be severely restricted by the condition of having eight
kinematical and eight dynamical supercharges and that of the
cancellation of nonabelian gauge anomalies of six-dimensional
worldvolume gauge theory as well as of the nonorientability of
the surface created by the Feynman diagrams. The rationales for the
choice of
the $usp$ Lie algebra and that of the field contents belonging to the
adjoint and antisymmetric representations have thus been given. The role
played by the degrees of freedom of the hypermultiplet in the
 fundamental representation remains, however, relatively unexplored.
 They participate in the anomaly cancellation mentioned above and
are responsible for creating an open string sector
as the counting of planar diagrams tells us.  That they do embody
$D3$ branes is less direct to grasp.

In this letter, we consider the $USp$ matrix model in the $T$ dualized
quantum mechanics where effects of these $D3$ branes could be seen as a
coupling of $D2$ magnetic background with the quantized degrees
of freedom of $D4$ branes. 
We find a residual interaction in the ``effective action'' for the
spacetime coordinates lying in the diagonal (Cartan) components of
the six adjoint matrices. This is accomplished by computing the Berry
 phase \cite{Berry,Stone}
coming from each of the quantum mechanics belonging to the three
fermionic sectors. We find an induced magnetic monopole background 
at a distance $m_{(f)}$ from the orientifold surface via the integrations
of fermions in the fundamental representation.
This is the effect which survives the cancellation of bosonic and
fermionic determinants.\footnote{ Papers dealing with related issues on
 fermions in matrix models include \cite{BD,ST,AIKKT}}.

\bigskip

{\bf Effective action for spacetime coordinates in the $USp$
matrix model:}

\bigskip

The starting point is the action of the zero dimensional $USp(2k)$
matrix model \cite{IT1,IT2}
\beqn 
 S= S_{{\rm vec}} + S_{{\rm asym}} + S_{{\rm fund}}\;\;.
\eeqn
 The part $S_{{\rm vec}} + S_{{\rm asym}}$  can be understood as
the projection from the type $IIB$ matrix model.  Introduce
a projector
\beqn
 \hat{\rho}_{\mp} \bullet  \equiv \frac{1}{2} \left( \bullet \mp F^{-1}
 \bullet^{t}  F \right) \;\;,
\eeqn
which  takes any $U(2k)$ matrix (denoted by a symbol with an underline)
into the matrix lying in the adjoint
representation  of $USp(2k)$  and that in the antisymmetric
representation respectively.  We obtain
\beqn
   S_{{\rm vec}} + S_{{\rm asym}} &=&  S_{{\cal N}=1}^{d=10}(
 \hat{\rho}_{b\mp}\ud{v}_{M}, \hat{\rho}_{f\mp}\ud{\Psi}  ) \;\;,
\label{eq:ours=10dSYM} \\
   S^{d=10}_{{\cal N}=1}(\ud{v}_{M},  \ud{\Psi} )  &=&
  \frac{1}{g^{2}} Tr \left( \frac{1}{4} \left[\ud{v}_{M},
 \ud{v}_{N} \right]
\left[ \ud{v}^{M}, \ud{v}^{N} \right] - \frac{1}{2}
\bar{\ud{\Psi}} \Gamma^{M} \left[ \ud{v}_{M},
\ud{\Psi} \right] \right) \;\;,
\eeqn
  where $\hat{\rho}_{b\mp} $ is a matrix with Lorentz indices
and  $\hat{\rho}_{f\mp} $ is a matrix with spinor indices:
\beqn
 \hat{\rho}_{b\mp}
 &=& diag
(\hat{\rho}_{-},\hat{\rho}_{-},\hat{\rho}_{-},
 \hat{\rho}_{-},\hat{\rho}_{-},\hat{\rho}_{+},
 \hat{\rho}_{+},\hat{\rho}_{-}, \hat{\rho}_{+},\hat{\rho}_{+} )
\nonumber \\
 \hat{\rho}_{f\mp}   &=&
\hat{\rho}_{-} 1_{(4)} \otimes
\left( \begin{array}{cccc}
        1_{(2)}& & & \\
               & 0 & & \\
               &   & 1_{(2)} & \\
               &   &         &0
        \end{array}
\right)
+
\hat{\rho}_{+} 1_{(4)} \otimes
\left( \begin{array}{cccc}
        0& & & \\
               & 1_{(2)} & & \\
               &   & 0 & \\
               &   &         & 1_{(2)}
        \end{array}
\right)
 \;\;.
\label{eq:projectors of b f}
\eeqn
 As for $S_{{\rm fund}}$, we work in the original representation based
on the four dimensional superfield notation with spacetime dependence
dropped:
\beqn
 S_{{\rm fund}} &=&  \frac{1}{g^{2}} \sum_{f=1}^{n_f}
\left[
  \int d^2 \th d^2 \thb
\left( Q_{(f)}^{* \;i} \left( e^{2V} \right)_i^{\;\;j} Q_{(f) \; j}
+  \tilde{Q}_{(f)}^{ i} \left( e^{-2V} \right)_i^{\;\;j}
        \tilde{Q}_{(f) \; j}^{*} \right)
\right.   \nonumber  \\
& &
\qquad
 +
\left.
\left\{
\int d^2 \th
\left(
        m_{(f)} \tilde{Q}_{(f)}^{\;\;\;\; i}  Q_{(f) \;i}
        + \sqrt{2} \tilde{Q}_{(f)}^{\;\;\;\; i}
                 \left( \Phi \right)_i^{\;\;j} Q_{(f) \;j}
\right)
+ h.c.
\right\}
\right]  \;\;, \\
Q_{i} &=&  Q_{i} + \sqrt{2} \th \psi_{Q \; i} + \th \th
 F_{Q \; i} \;\;.
\eeqn
  For more complete definition of the action,  see \cite{IT1,IT2}.
The mass term which we have denoted by $m_{(f)}$ is necessary for the  
discussion in what follows.
In the leading large $k$ (planar) limit in the sense  of 't Hooft,
$S_{{\rm fund}}$ is ignorable.  Any physical consequence coming from
$S_{{\rm fund}}$ must be from processes  which receive a vanishing  
contribution from the planar diagrams.

Physical quantities of the $USp(2k)$ reduced model are obtained
from the effective action for the spacetime coordinates $x_{M}$,
( which  are the diagonal elements of $v_{M}$),
\beqn
Z \left[ x_{M} ; m_{(f)}\right]
&=&
\int \left[ D \tilde{v}_{M} \right]
\left[ D \Psi\right] \left[ D \overline{\Psi} \right]  
\prod_{f=1}^{n_{f}}
\left[ D Q_{(f)}\right] \left[ D Q_{(f)}^{*}\right]
\left[ D {\tilde{Q}}_{(f)} \right]
\left[ D \tilde{Q}_{(f)}^{*} \right]
\left[ D \psi_{Q (f)}    \right] \nonumber  \\
&\times&
\left[ D  {\overline{\psi}}_{Q (f)}   \right]
\left[ D  \psi_{{\tilde{Q}} (f)}    \right]
\left[ D  {\overline{\psi}}_{{\tilde{Q}} (f)} \right]
\exp[iS] \;\;\; \nonumber  \\
&=&
\int \left[ D \tilde{v}_{M} \right]
\prod_{f=1}^{n_{f}}
\left[ D Q_{(f)}\right] \left[ D Q_{(f)}^{*}\right]
  \left[D {\tilde{Q}}_{(f)} \right]
\left[ D \tilde{Q}_{(f)}^{*} \right] \nonumber  \\
&\times&
\left[ \det D_{{\rm fund}}(v_{M})\right]^{n_{f}}
\left[ \det D_{{\rm adj}}(v_{M})\right]
\left[ \det D_{{\rm asym}}(v_{M})\right]
\exp[iS_{B}] \;\;\;
\eeqn
and all possible operator insertions ( local as well as
nonlocal ones) into this object.
  Here  $v_{M} = x_{M}+ \tilde{v}_{M}$ and  $S^{B}=S$ with all
 fermions set to zero.  For simplicity in this paper we keep only
 the diagonal elements of  six adjoint directions
\beqn
x_{\nu} =
 diag(x_{\nu}^{(1)}, \cdots x_{\nu}^{(k)}, -x_{\nu}^{(1)},
 \cdots -x_{\nu}^{(k)}) \;\;.
\eeqn
  At one-loop, it is legitimate to replace  the matrix
 $v_{\nu}= x_{\nu}+ \tilde{v}_{\nu}$ in the argument of
the determinant by the diagonal matrices  $x_{\nu}$.

 Naively supersymmetry would tell the cancellation of the bosonic
determinants against the fermionic ones.  As this can be done by  
the $1PI$ Feynman diagrams, the cancellation, if true, would persist  
to all orders in perturbation theory. To see that this is not quite
the case, we  $T$ dualize the system.
 Recall that the $T$ duality transformation is a legitimate
operation in the large $k$ limit via the recipe of \cite{WT}.
We regard the fermionic integrations with a particular
set of nonlocal operator insertions as the transition
amplitude of an adiabatic process.  This latter process is given
by the quantum mechanical systems of free fermions with external
bosonic parameters $x_{\nu}$ and $m_{(f)}$.
The object which we will study in what follows is
\beqn
\label{eq:adprocess}
Z \left[ x_{\nu}, x_{I}=0 ; m_{(f)}; j_{(R)} \right] =
  \int \left[ D \tilde{v}_{M} \right]    \prod_{f=1}^{n_{f}}
\left[ D Q_{(f)}\right] \left[ D Q_{(f)}^{*}\right]
  \left[D {\tilde{Q}}_{(f)} \right]
\left[ D \tilde{Q}_{(f)}^{*} \right]  \exp [ iS_{B} ]  \nonumber \\
 \lim_{T \rightarrow \infty}
 \prod_{f=1}^{n_{f}} \left[\langle t=T; j_{f} \mid P
 e^{ -i \int_{0}^{T}
 dt H_{{\rm fund}}(t)} \mid t=0;j_{f} \rangle^{(f)} \right]   \\
  \langle t=T ;  j_{{\rm adj}} \mid P e^{-i \int_{0}^{T}
 dt H_{{\rm adj}}(t)} \mid t=0; j_{{\rm adj}} \rangle
 \langle t=T; j_{{\rm asym}} \mid P e^{ -i \int_{0}^{T}
 dt H_{{\rm asym}}(t)} \mid t=0 ; j_{{\rm asym}} \rangle
 \;\;\;.  \nonumber 
\eeqn
 Here we denote by $H_{{\rm fund}}(t), H_{{\rm adj}}(t),
 H_{{\rm asym}}(t),$
the respective Hamiltonians obtained from the fermionic part of
$ S_{{\rm fund}}, S_{{\rm adj}}$ and $S_{{\rm asym}}$ after $T$ duality.
Their $t$ dependence  comes from  that of the diagonal matrix  
$x_{\ell}$  which act as external parameters on the Hilbert space
of fermions. We have indicated by
$j_{(R)}$   ( $R = {\rm fund, adj, antisym}$ )
quantum numbers of an adiabatic state generically.
Let us denote by $e_{(R)}^{(A)}$ the standard eigenbases
belonging  to the roots of $sp(2k)$ and the weights of
the fundamental representation and those of the antisymmetric
representation respectively.
We expand the two component fermions as
\beqn
\label{eq:expand}
 \psi^{(R)} = \sum_{A}^{N_{(R)}} b_{A}^{(R)}
 e_{(R)}^{(A)} /\sqrt{2} \;,\;\;
\bar{\psi}^{(R)} = \sum_{A}^{N_{(R)}} \bar{b}_{A}^{(R)}
 e_{(R)}^{(A) \dagger}/ \sqrt{2} \;\;\;,
\eeqn
  where  $N_{({\rm adj})}= 2k^{2} +k$,
  $N_{({\rm antisym})}= 2k^{2} -k$ and  $N_{({\rm fund})}= 2k$.
  We find that all of the three Hamiltonians
 $H_{{\rm fund}},H_{{\rm adj}}$ and $H_{{\rm asym}}$
 are expressible in terms of the abelian counterpart
\beqn
\label{eq:abelian}
  g^{2} H_{0} \left(x_{\ell}, \phi, \phi^{*}; (R), A \right) =
-\bar{b}_{A\dot{\a}}^{(R)} {\sb}^{m\dot{\a}\a}x_{m}b_{A\a}^{(R)}
-d^{(R)\a}_{A} \s_{\a\dot{\a}}^{m}x_{m} \bar{d}_{A}^{(R) \dot{\a}}
+ \sqrt{2}  \phi b^{(R) \a}_{A}  d_{A\a}^{(R)}  \nonumber \\
+\sqrt{2}\phi^{*}\bar{b}_{A\dot{\a}}^{(R)}
  \bar{d}^{\dot{\a} (R) }_{A} \;\;
\eeqn
provided  we replace the five parameters
\beq
x_{\ell},\;\;\; \phi = \frac{x_4 + ix_7}{\sqrt{2}}, \;\;\;
\phi^{*} = \frac{x_4 - ix_7}{\sqrt{2}}
\eeq
by the appropriate ones. (See argument of $\gamma_{\Gamma}$ in  
eq.~(\ref{eq:gamma}) below.)
  
 The formula for the transition amplitude of an adiabatic process is
\beqn
 \lim_{T \rightarrow \infty} \langle t= T; j \mid P \exp \left[ -i  
\int_{0}^{T}
 dt H_{0}(t) \right] \mid t=0; j \rangle
  &=& \exp \left[ -i \int_{0}^{T} E(t) dt
 + i \gamma_{\Gamma} \right]\;\;, \nonumber \\
  \gamma_{\Gamma} \left[ x_{m}, \phi, \phi^{*}; j \right]  &=&
 \int_{0}^{T} dt
 \frac{d\gamma }{dt} \;\;\;.
\eeqn
Here $\Gamma$ is a closed path in the parameter space. The  
connection one-form associated with the Berry phase  
$\gamma_{\Gamma}$ satisfies
\beqn
  id\gamma (t)= -  \langle t \mid d \mid t\rangle
  \equiv   -i {\cal A} \;\;\;.
\eeqn
Using this expression, the second and the third lines of
 eq.~(\ref{eq:adprocess}) are written as
\beqn
\label{eq:gamma}
\exp (i \gamma_{\Gamma}^{(total)}) \equiv
\exp \left( i \sum_{f=1}^{n_{f}} \sum_{A=1}^{2k}
\gamma_{\Gamma}
\left[ {\bf w}^{A}\cdot {\bf x} _{\ell},\;
  m_{(f)}/\sqrt{2}    +{\bf w}^{A} \cdot {\bf\Phi},\;
  m_{(f)}/\sqrt{2} + {\bf w}^{A} \cdot {\bf \Phi }^{\dagger}
 \right] \right.  \nonumber \\
\left.  +  i \sum_{A=1}^{2k^{2}}
\gamma_{\Gamma} \left[ {\bf R}^{A} \cdot {\bf x}_{\ell},\;
 i {\bf R}^{A} \cdot {\bf \Phi},\;
  i {\bf R}^{A} \cdot {\bf \Phi}^{\dagger} \right]
 +i  \sum_{A=1}^{2k^2 - 2k}
\gamma_{\Gamma} \left[ {\bf w}_{{\rm asym}}^{A} \cdot {\bf x}_{\ell},\;
 {\bf w}_{{\rm asym}}^{A} \cdot {\bf \Phi},\;
  {\bf w}_{{\rm asym}}^{A} \cdot {\bf \Phi}^{\dagger} \right] \right).
\eeqn
Here
\beqn
\{ \{  {\bf w}^{A} \mid  1 \leq A \leq 2k  \}\}
&=&
\{ \{   \pm {\bf e}^{(i)}
\; , 1 \leq i \leq k   \}\}   \;\; {\rm and} \nonumber \\
\{ \{ {\bf R}^{A} \mid 1 \leq A \leq 2k^{2} \}\}
&=&
\{ \{   \pm 2 {\bf e}^{(i)}, {\bf e}^{(i)}-{\bf e}^{(j)} ,
\pm \left({\bf e}^{(i)} +{\bf  e}^{(j)} \right)
\; 1 \leq i,j, \leq k   \}\}  \;\;{\rm and} \nonumber \\
\{ \{  {\bf w}_{ {\rm asym} }^{A} \mid  1 \leq A \leq 2k^{2} -2k   
\}\}
&=&
\{ \{    \pm \left( {\bf e}^{(i)} +{\bf  e}^{(j)}  \right) ,
{\bf e}^{(i)} - {\bf e}^{(j)},  \; 1 \leq i,j, \leq k   \}\}
\eeqn	
 are respectively the nonzero roots and the weights in the  
antisymmetric representation
 of $usp(2k)$. We have denoted by ${\bf e}^{(i)} \;(1 \leq i \leq  k)$
the orthonormal basis vectors of $k$-dimensional Euclidean space and
\beqn
{\bf x}_{\ell} = \sum_{i=1}^{k}  {\bf e}^{(i)} x_{\ell}^{(i)},\;
{\bf \Phi} =  \sum_{i=1}^{k}{\bf e}^{(i)}
\frac{x_{4}^{(i)} + ix_{7}^{(i)}}{\sqrt{2}}, \;
{\bf \Phi}^{\dagger}= \sum_{i=1}^{k}{\bf e}^{(i)}
\frac{x_{4}^{(i)} - ix_{7}^{(i)}}{\sqrt{2}}
\;\;\;.
\eeqn

\bigskip
 
{\bf Computation of the Berry connection:}

\bigskip

   Let us now turn to the computation of $\gamma_{\Gamma}(t)$
 associated with $H_{0}( x_{\ell}, \phi, \phi^{*} )$.
 We define the Clifford vacuum $\ket{\Omega}$ by
\beq
b^{\a}\ket{\Omega} = \bar{d}_{\dot{\a}}\ket{\Omega} = 0.
\eeq
  We have suppressed the labels $A$ and $(R)$ seen in
 eqs.(\ref{eq:expand}),(\ref{eq:abelian}).
 Any ket vector of this system can be decomposed into a set of wave  
functions by
\beqn	
\ket{\;\;\;}
=
\left[
  h_{(4)}\frac{1}{4} \bar{b}_{\dot{\a}}\bar{b}^{\dot{\a}}
d^{\a}d_{\a}
+ h_{(3)\a} \half d^{\a} \bar{b}_{\dot{\a}}\bar{b}^{\dot{\a}}
+ \bar{h}_{(3)}^{\;\;\dot{\a}} \half \bar{b}_{\dot{\a}}
d^{\a} d_{\a}
+ h_{(2,1)} \half \bar{b}_{\dot{\a}} \bar{b}^{\dot{\a}}
+ h_{(2,2)} \half d^{\a}d_{\a}
\nonumber \right. \\ \left.
+ h_{(2,3)} \half d^{\a} \s_{\a \dot{\a}}^0 \bar{b}^{\dot{\a}}
+ h_{(2,4)i} \half d^{\a} \s_{\a \dot{\a}}^i \bar{b}^{\dot{\a}}
+ h_{(1)\a} d^{\a}
+ \bar{h}_{(1)}^{\dot{\a}} \bar{b}_{\dot{\a}}
+ h_{(0)}
\right]
\ket{\Omega},
\eeqn
  As the ``particle number'', which we denote by  $n$,  is  
conserved,
 the eigenvalue problem reduces
 to those in each sector $n= 0,1,2,3,4$.  Both of the
 $n=0,4$  sectors give a zero eigenvalue trivially  while
 $n=1$  and $n=3$ sectors are related to each other by
  $b^{\a} \leftrightarrow \bar{b}_{\dot{\a}}, \;\;
  \bar{d}_{\dot{\a}}  \leftrightarrow  d^{\a}$.
  We are left to analyze
\beqn
\label{eq:MM}
M_{3}
\left(
\begin{array}{c}
 h_{(3)\alpha} \\
 \bar{h}_{(3)}^{\;\;\dot{\alpha}}
\end{array}
\right)
 =
g^{2}  E
\left(
\begin{array}{c}
 h_{(3)\alpha} \\
 \bar{h}_{(3)}^{\;\;\dot{\alpha}}
\end{array}
\right), \;\;  {\rm and \;\;} \;
M_2
\left(
\begin{array}{c}
 h_{(2,1)} \\
 h_{(2,2)} \\
 h_{(2,3)} \\
 h_{(2,4)i}
\end{array}
\right)
=   g^{2}
E
\left(
\begin{array}{c}
 h_{(2,1)} \\
 h_{(2,2)} \\
 h_{(2,3)} \\
 h_{(2,4)i}
\end{array}
\right)\;\;.
\eeqn
Here $M_{3}$ and $M_{2}$ are representation matrices
 for $g^{2}H_{0}$  of an appropriate size :
\beqn
M_3  &=&
\left(
\begin{array}{cccc}
-x_3 & -x_1 - ix_2 &  \sqrt{2} \phi^{*} & 0      \\
-x_1 + ix_2 & x_3 & 0 &  \sqrt{2} \phi^{*} \\
 \sqrt{2} \phi & 0 & x_3 & x_1+ ix_2        \\
0 &  \sqrt{2} \phi & x_1 - ix_2 & -x_3
\end{array}
\right) \;\;.  \\
M_2 &=&
\left(
\begin{array}{cccc}
0 & 0 &  \sqrt{2} \phi^{*} & 0      \\
0 & 0 & - \sqrt{2} \phi & 0 \\
 \sqrt{2} \phi & - \sqrt{2} \phi^{*} & 0 & -2x^i
\\
0 & 0 & -2x_i & 0
\end{array}
\right) \;\;.
\eeqn
The eigenvalues of $M_{3}$ are
\beq
 \pm \sqrt{ x_{\ell}x^{\ell} +
2\phi \phi^{*}  }  \equiv \pm \lambda_{0}  \;\;,
\eeq
 each being two-fold degenerate. The eigenvalues of $M_{2}$ are
\beq
 0,  \;\; \pm 2 \lambda_{0}\;\;,
\eeq
 where the zero eigenvalue is four-fold degenerate.   Alternatively
 one can show these by constructing the eigenmode operators of  
$H_{0}$.
 
  We now analyze the bundle structure associated with the first one  
of
 eq.~(\ref{eq:MM}).
  Let us write
\beqn
\left(
\begin{array}{c}
 h_{(3)\alpha} \\
 \bar{h}_{(3)}^{\;\;\dot{\alpha}}
\end{array}
\right)
  \equiv  {\bf h}_{A, a}   \;\;\;,
\eeqn
 where the first subscript of the right hand side refers to  
$h_{(3)}$ or
 $\bar{h}_{(3)}$ and the second one to the spinor indices.
 Introducing five dimensional spherical coordinates
\beqn
x_2 &=& r \sin \phi_1 \sin \theta_1 \cos \theta_2  \;\;\;, \nonumber  
\\
x_1 &=& r \cos \phi_1 \sin \theta_1 \cos \theta_2 \;\;\;,  \nonumber  
\\
x_3 &=& r  \;\;\;\;\;\;\;\;\;     \cos \theta_1 \cos \theta_2
 \;\;\;, \nonumber \\
x_4  &=& r \;\;\;\;\;\;\;\;\;\;\;\;\;\;\;\;\;\;
 \sin \theta_2 \cos \phi_2  \;\;\;, \;\; \qquad 0 \leq \phi_2 \leq 2  
 \pi \;\;, \nonumber \\
x_7 &=& r   \;\;\;\;\;\;\;\;\;\;\;\;\;\;\;\;\;\;
 \sin \theta_2 \sin \phi_2  \;\;\;, \;\; \qquad 0 \leq \theta_2 \leq  
\pi \;\;,
\eeqn
 we  find
\beq
{\cal M} (\theta_2 , \phi_2 ,\theta_1 , \phi_1)  \equiv \frac{1}{r}  
M_3 =
\left(
\begin{array}{cc}
- \cos{\theta_2} {\bf U}( \theta_1 , \phi_1 ), & \sin{\theta_2}
 e^{-i\phi_2} {\bf 1}_{(2)}  \\
\sin{\theta_2} e^{i\phi_2} {\bf 1}_{(2)},  & \cos{\theta_2}
{\bf U}( \theta_1 , \phi_1 )
\end{array}
\right),
\eeq
where
\beq
{\bf U}( \theta_1 , \phi_1 ) =
\left(
\begin{array}{cc}
\cos{\theta_1} & \sin{\theta_1} e^{i\phi_1} \\
\sin{\theta_1} e^{-i\phi_1} & - \cos{\theta_1}
\end{array}
\right).
\eeq
Note that
\beq
{\cal M} (\theta_2 , \phi_2 ,\theta_1 , \phi_1)^2 =
 {\bf 1}_{(2)} \otimes {\bf 1}_{(2)}
 \equiv {\bf 1}_{4}
 , \quad
{\bf U}( \theta_1 , \phi_1 )^2 = {\bf 1}_{(2)},
\eeq
and
\beq
[ \;{\cal M} (\theta_2 , \phi_2 ,\theta_1 , \phi_1),
\; {\bf 1}_{(2)} \otimes   {\bf U}( \theta_1 , \phi_1 ) \;]
= 0 \;\;.
\eeq
  The four eigenstates are specified by the
eigenvalues $\pm 1$ of $ {\cal M} (\theta_2 , \phi_2 ,\theta_1 ,  
\phi_1) $
 and  those of $ {\bf 1}_{(2)} \otimes {\bf U}( \theta_1 , \phi_1  
)$,
  which are obtained by the projection operators
\beq
 \half \left(  {\bf 1}_{(4)} \pm {\cal M} \right) \half \left(  {\bf  
1}_{(4)}
 \pm   {\bf 1}_{(2)} \otimes  {\bf U} \right)  {\bf h}_{A, a} \;\;
\eeq
  up to a numerical constant.

  From now on, we focus on the $(+, +)$ case without losing  
generality.
The normalized eigenfunction is actually a section. Indicating its  
local forms
 around $\left( \theta_{2},\theta_{1} \right)$
$= \left(0, 0 \right), \left(0, \pi \right),$ 
 $\left( \pi, 0 \right),$ and  $\left( \pi, \pi \right)$ by
$(N,N),(N,S),(S,N)$ and $(S,S)$  respectively, we find
\beqn
\label{eq:efn}
{\bf h}_{A, a}^{(N,N)}
&=&
\left(
\begin{array}{c}
 \sin \frac{\theta_2}{2} \, e^{-i \phi_2} \\
 \cos \frac{\theta_2}{2} \\
\end{array}
\right)_{A}
\otimes
\left(
\begin{array}{c}
 \cos \frac{\theta_1}{2} \\
 \sin \frac{\theta_1}{2} \, e^{i \phi_1} \\
\end{array}
\right)_{a}\;\;,
\nonumber \\
{\bf h}_{A, a}^{(N,S)}
&=&
\left(
\begin{array}{c}
 \sin \frac{\theta_2}{2} \, e^{-i \phi_2} \\
 \cos \frac{\theta_2}{2} \\
\end{array}
\right)_{A}
\otimes
\left(
\begin{array}{c}
 \cos \frac{\theta_1}{2} \, e^{-i \phi_1} \\
 \sin \frac{\theta_1}{2} \\
\end{array}
\right)_{a}  \;\;,
\nonumber \\
{\bf h}_{A, a}^{(S,N)}
&=&
\left(
\begin{array}{c}
 \sin \frac{\theta_2}{2} \\
 \cos \frac{\theta_2}{2}  \, e^{i \phi_2} \; \\
\end{array}
\right)_{A}
\otimes
\left(
\begin{array}{c}
 \cos \frac{\theta_1}{2} \\
 \sin \frac{\theta_1}{2} \, e^{i \phi_1} \\
\end{array}
\right)_{a} \;\; ,
\nonumber \\
 {\bf h}_{A, a}^{(S,S)}
&=&
\left(
\begin{array}{c}
 \sin \frac{\theta_2}{2}  \\
 \cos \frac{\theta_2}{2}  \, e^{i \phi_2} \; \\
\end{array}
\right)_{A}
\otimes
\left(
\begin{array}{c}
 \cos \frac{\theta_1}{2} \, e^{-i \phi_1} \\
 \sin \frac{\theta_1}{2} \\
\end{array}
\right)_{a}   \;\;\;.
\eeqn

  This is obviously a simplest generalization of the original  
problem discussed in \cite{Berry}. ( See also \cite{Stone}.)
  We obtain the connection one-form
\beqn
\label{eq:BC}
{\cal{A}}^{(N,N)} &=&
- \frac{i}{2} ( 1 - \cos \theta_2 ) d \phi_2
 + \frac{i}{2} ( 1 - \cos \theta_1 ) d \phi_1   \;\;,
\nonumber \\
{\cal{A}}^{(N,S)} &=&
- \frac{i}{2} ( 1 - \cos \theta_2 ) d \phi_2
 - \frac{i}{2} ( 1 + \cos \theta_1 ) d \phi_1 \;\;,
\nonumber \\
{\cal{A}}^{(S,N)} &=&
+\frac{i}{2} ( 1 + \cos \theta_2 ) d \phi_2
+ \frac{i}{2} ( 1 - \cos \theta_1 ) d \phi_1 \;\;,
\nonumber \\
{\cal{A}}^{(S,S)} &=&
+ \frac{i}{2} ( 1 + \cos \theta_2 ) d \phi_2
- \frac{i}{2} ( 1 + \cos \theta_1 ) d \phi_1 \;\;,
\eeqn

  We have also determined the normalized eigenfunctions for
the second eq. of (\ref{eq:MM}), namely, the
two-particle case, which  we now describe only briefly and  
qualitatively. In contrast to eq. (\ref{eq:efn}),  the eigenfunctions
belonging to zero or $\pm 2 \lambda_{0}$  are described by
ordinary functions, not developing into expressions involving
half angles indicative of singularities.  For the nonzero
eigenvalues, we find the vanishing connection while the states with
zero eigenvalue give a pure gauge configuration and  are gauged
away.  We conclude that
the $n=1, 3$ sectors, or equivalently,  the first and  the third  
excited states give rise to the nontrivial connection of
the form of eq. (\ref{eq:BC}) in the parameter space while the
remaining states including the ground state
(with $\lambda_{0} = -2$) do not.

\bigskip

{\bf Brane interpretation:}
\bigskip

Let us now apply the formula (eq.~(\ref{eq:BC})) we obtained to our
original problem.  Due to the symmetry of the roots and the weights
under ${\bf e}^{(i)} \leftrightarrow -{\bf e}^{(i)}$, the contributions
from the adjoint and antisymmetric representations (the second and
the third  terms in the exponent of eq.~(\ref{eq:gamma})) cancel
when summed over $A$.
The cancellation occurs as well to the part from the fundamental  
representation which does not involve $\phi^{(i)}$ or $\phi^{(i)*}$.
We find that $\gamma_{\Gamma}^{({\rm total})}$ in
eq.~(\ref{eq:gamma}) is
written as
\beqn
   \gamma_{\Gamma}^{({\rm total})} &=&
 \sum_{f=1}^{n_{f}} \sum_{i=1}^{k} 
 \gamma_{\Gamma}^{({\rm Berry})}\left[  x_{3}^{\prime (i)},
  m_{f}/\sqrt{2}+ \phi^{(i)}, m_{(f)}/\sqrt{2} + \phi^{(i)*} \right]
    \nonumber \\
 &+& \sum_{f=1}^{n_{f}} \sum_{i=1}^{k}
 \gamma_{\Gamma}^{({\rm Berry})}\left[  x_{3}^{\prime (i)},
m_{f}/\sqrt{2}- \phi^{(i)}, m_{(f)}/\sqrt{2} - \phi^{(i)*}
\right] \;\;.
\eeqn
Here
\beqn
\gamma_{\Gamma}^{({\rm Berry})} \left[ x_{3}^{\prime}, \phi, \phi^{*}
\right]
  &=&  \int {\cal A}^{({\rm Berry})} \\  
{\cal{A}}^{(N)({\rm Berry})} &=&
- \frac{i}{2} ( 1 - \cos \theta_2 ) d \phi_2 \;, \;\;
{\cal{A}}^{(S)({\rm Berry})} =
+\frac{i}{2} ( 1 + \cos \theta_2 ) d \phi_2 \;\;. \label{eq:config}
\eeqn
and $x_{3}^{\prime (i) 2} = x_{1}^{(i)2} + x_{2}^{(i)2}+ x_{3}^{(i)2}$.

    It is satisfying to see a pair of magnetic monopoles sitting at
$x_{4}^{(i)} = \pm m_{(f)}$ from the orientifold surface
for $i= 1 \sim k$.  ( See \cite{IT2} for the presence of orientifold
surfaces in the $USp$ matrix model.) These monopoles live in the
parameter space,
which is the spacetime coordinates generated by the matrix model.
Coming back to eq.~(\ref{eq:adprocess}), we conclude that the Berry
phase generates an interaction
\beqn
\label{eq:induced}
Z \left[ x_{\ell}, x_{I}=0 ; \cdots\right] =
  \int \left[ D \tilde{v}_{M} \right]    \prod_{f=1}^{n_{f}}
\left[ D Q_{(f)}\right] \left[ D Q_{(f)}^{*}\right]
  \left[D {\tilde{Q}}_{(f)} \right]
\left[ D \tilde{Q}_{(f)}^{*} \right]  \exp [ iS_{B} +  
 i \gamma_{\Gamma}^{({\rm total})} ]  \;\;.
\eeqn

Let us give this configuration we have obtained a brane
interpretation \cite{Pol} first from the six dimensional and
subsequently from the ten dimensional
point of view. It should be noted that the two coordinates
which the connection ${\cal{A}}^{({\rm Berry})}$ does not depend
on are the angular cooordinates $\theta_{1}, \phi_{1}$, so that 
$x_{1}, x_{2}$ are not quite separable from the rest of the
coordinates $x_{3}, x_{4}, x_{7}$  in
eq.~(\ref{eq:config}).  Only in the asymptotic region 
$\mid x_{3}^{\prime} \mid >> \mid x_{3} \mid$, there exists
an area of size $\pi \mid x_{3}^{\prime} \mid^{2}$ transverse to
the three dimensional space  where the Berry phase is obtained.
In this region, the magnetic flux obtained from the $b(=1)$-form
connection embedded in $d(=6)$-dimensional spacetime looks
approximately as is discussed in \cite{Nepomechie}: the flux
no longer looks coming from a poinlike object but from
a $d-b-3(=2)$ dimensionally extended object.
The magnetic monopole obeying the Dirac quantization behaves
approximately like a magnetic $D2$ brane \cite{Pol} extending to the
(1,2) directions, which are perpendicular to the orientifold surface.
In fact, the presence of this object and its quantized magnetic flux
have been detected by quantum mechanics of a point particle (electric
$D0$ brane) obtained from the $n=1$ and $n=3$ particle states of
the fermionic sector in the fundamental  representation.
The induced interaction is a minimal one. We conclude that the $D0$  
represented by the first and the third excited states of the quantum  
mechanical problem given above is under the magnetic field created
by $D2$. To include the four remaining coordinates
$(x_5, x_6, x_8, x_9)$ of the  antisymmetric directions, we appeal
to the translational invariance which is preserved in these
directions.  The simplest possibility is that they appear in the
coupling through the derivatives
\beqn
\int {\cal A}^{({\rm Berry})} =
\int\prod_{I=5,6,8,9} dX^{I}dX^{\nu} A_{\nu 5689}  \;\;.
\eeqn 
With this assumption,\footnote{ Similar reasoning is seen in \cite{BD}.}
the $D0$ brane is actually a $D4$ bane extended in $(5,6,8,9)$
directions while the magnetic $D2$ still occupies $(1,2)$: the
quantization condition is preserved in ten dimensions as well.

 We have exhibited a residual interaction due to the fermions in the
fundamental representation. In the situation we have dealt with
as quantum mechanics, however, the magnetic flux escapes to infinity.
As a result, there is no conservation law which fixes the number of
flavors $n_{f}$.  Let us finally make a brief comment on the case in
which we compactify all six adjoint directions. As is shown in 
\cite{IT2}, the number of flavour is determined to be $n_{f} =16$ by
the cancellation of the nonabelian anomaly of $6d$ worldvolume
gauge theory obtained from the $USp$ matrix model via the recipe
of \cite{WT}. We can apply to this case the same line of treatment
given above based on the Berry phase. In fact, following the works of
ref. \cite{anomaly}, we find that the residual interaction is given
by the two-parameter integral of the anomalous commutator
\footnote{ For other applications of anomalies and anomalous
interactions to $D$ branes, see, for example, \cite{brane}.}  of
the Gauss law generators.
This is a two cocycle $w_{2}$ of the gauge group on $R^{5}$ obtained
from
\beq
 w_{-1} = \frac{1}{4 ! (4\pi i)^{4}} Str \left( F \wedge F \wedge
 F \wedge F  \right)
\eeq
 via descent equations
\beq
 w_{-1} = dw_{0}\;, \;\; \delta w_{0} = dw_{1}\;, \;\;
 \delta w_{1} = dw_{2}\;\;.
\eeq
The cancellation of the strength of the anomalous commutator fixes
$n_{f}=16$.

\bigskip

 The authors thank Yasuhito Arakane, Asato Tsuchiya and Takashi Yokono
 for helpful discussion on this subject.


\newpage

\begin{thebibliography}{99}

\bibitem{BFSS}
T. Banks, W. Fischler, S. Shenker and L. Susskind,
{\sl Phys. Rev.} {\bf D55 }(1997) 5112, hep-th/9610043.

\bibitem{Mhetero}
N. Kim and S.J. Rey, hep-th/9701139;
T. Banks and L. Motl, hep-th/9703218;
D.A. Lowe, hep-th/9704041;
S.J. Rey, hep-th/9704158;
P. Horava, hep-th/9705055;
M. Krogh, hep-th/9801034, hep-th/9803088

\bibitem{IKKT}
N. Ishibashi, H. Kawai, Y. Kitazawa and A. Tsuchiya,
{\sl Nucl. Phys.} {\bf  B498 }(1997) 467.

\bibitem{IT1}
H. Itoyama and A. Tokura, {\sl Prog. Theor. Phys.} {\bf 99} (1998)129,
hep-th/9708123.

\bibitem{Vafa}
 C. Vafa, {\sl Nucl.Phys.}{\bf B469}(1996)403.

\bibitem{IT2}
H. Itoyama and A. Tokura, {\sl Phys. Rev.D} {\bf D58} (1998)026002,
 hep-th/9801084.

\bibitem{Berry}
M. V. Berry, {\sl Proc. R. Soc. Lond.} {\bf A392} (1984)45.

\bibitem{Stone}
M. Stone, {\sl Phys. Rev.} {\bf D33} (1986)1191.

\bibitem{BD}
M. Berkooz and M. R. Douglas,
{\sl Phys. Lett.} {\bf B395} (1997)196.

\bibitem{ST}
T. Suyama, and A. Tsuchiya, {\sl Prog. Theor. Phys. 99} (1997) 321  
hep-th/9711073.

\bibitem{AIKKT}
H. Aoki, S. Iso, H. Kawai, Y. Kitazawa and T. Tada, hep-th/9802085.

 
\bibitem{WT}
W. Taylor IV, {\sl Phys. Lett.} {\bf B394} (1997) 283  
hep-th/9611042.
\\
O.J. Ganor, S. Ramgoolam and W. Taylor IV,
{\sl Nucl. Phys.} {\bf B492} (1997) 191 hep-th/9611202.

\bibitem{Pol}
J. Polchinski,
{\sl Phys. Rev. Lett.} {\bf 75} (1995)4724.
\\
J. Polchinski, {\it TASI Lectures on D-branes \/} 
hep-th/9611050.

\bibitem{Nepomechie}
R.I. Nepomechie, {\sl Phys. Rev.} {\bf D31} (1985)1921.
\\
C. Teitelboim,
{\sl Phys. Lett.} {\bf 167B} (1986)63, 69.

\bibitem{anomaly}
L. D. Faddeev, {\sl Phys. Lett} {\bf 145B} (1984) 81.
\\
P. Nelson and L. Alvarez-Gaum$\acute{\mbox{e}}$,
{\sl Comm. Math. Phys.} {\bf99} (1985)103.
\\
H. Sonoda,
{\sl Phys. Lett.} {\bf 156B} (1985)220.

\bibitem{brane}
M. Green, J. A. Harvey and G. Moore, {\sl Class. Quant. Grav. 14}  
(1997)47 hep-th/9605033.
\\
C. Bachas, M. Douglas and M. Green, hep-th/9705074.

\end{thebibliography}
\end{document}